# Ingot Laser Guide Stars Wavefront Sensing

Roberto Ragazzoni[a,b], Elisa Portaluri[a,b], Valentina Viotto[a,b], Marco Dima[a,b],
Maria Bergomi[a,b], Federico Biondi[a,c], Jacopo Farinato[a,b], Elena Carolo[a,b],
Simonetta Chinellato[a,b], Davide Greggio[a,c], Marco Gullieuszik[a],
Demetrio Magrin[a,b], Luca Marafatto[a,b], Daniele Vassallo[a,b,c]

[a] INAF - Osservatorio Astronomico di Padova, Vicolo dell'Osservatorio 5, 35122 Padova, Italy
[b] ADONI – Laboratorio nazionale per l'Ottica Adattiva, Italy
[c] Dipartimento di Fisica ed Astronomia - Università degli Studi di Padova,
Vicolo dell'Osservatorio 3, 35122 Padova, Italy

## ABSTRACT

We revisit one class of z-invariant WaveFront sensor where the LGS is fired aside of the telescope aperture. In this way there is a spatial dependence on the focal plane with respect to the height where the resonant scattering occurs. We revise the basic parameters involving the geometry and we propose various merit functions to define how much improvement can be attained by a z-invariant approach. We show that refractive approaches are not viable and we discuss several solutions involving reflective ones in what has been nicknamed "ingot wavefront sensor" discussing the degrees of freedom required to keep tracking and the basic recipe for the optical design.

## 1. INTRODUCTION

Laser Guide Stars (LGS hereafter) have been introduced in order to produce artificial reference sources for Adaptive Optics compensation[1]. A thorough analysis of the various possible combinations of wavelength and excitation of layers in the atmosphere[2] left, basically two options: one to use the Rayleigh scattering from the lower portion of the atmosphere, a phenomenon basically independent by any specific wavelength but much more effective toward the shorter ones, and the mesospheric resonant scattering by a natural layer of Sodium. The latter are often referred to as Sodium LGSs. While most of the development in the previous years have been devoted to achieve a stable, reliable, and economically effective source of laser light with the right characteristics, the way the LGS is fired in the sky has been the subject of a sort of natural development. As these relatively expensive devices are likely to occurs first on large facilities, multiple LGSs are mandatory and firing them from the side of the telescope sounds to be the obvious natural choice. Early concepts[3] pointed out that such LGSs are not representative of a "star" at all but in fact are sort of "cigars in the sky" that exhibit simultaneously an apparent elongation, as seen from aside, and a continuous amount of defocusing, as the source of radiation span more than a decade of km over a range that is just one order of magnitude larger. Later these crude concept evolved into a class of WFS that has been nicknamed "z-invariant"[4,5,6] recalling the fact that they deploy along an axis close to the optical one (usually denoted by z) and has been tested in the sky with Rayleigh beacon[7,8] and potential extensions to the realm of ELTs have been described[9].

We resume here the original concept described so far in[3] and we try to generalize and to give a first rough estimation of the quantities involved and the figures that should characterize their potential advantage upon other non-moving parts approach.

## 2. ELONGATION IN A 3D WORLD

Sodium LGSs are extended sources. Once projected from aside each point along the intersection of the LGS projector line of sight with the Sodium layer will be seen at a different angular displacement and at a different defocus terms (or range of conjugation) and will translates into a focusing in the volume behind the conventional, starlight, focal plane into different points. In Fig.1 an exemplification is shown offering three different examples. One should remind that the view depicted here is just a crude exemplification. In the true world LGSs can be projected with a propagation direction that is tilted with respect to the telescope optical axis, or that it can be even skewed. Furthermore the position of the exit pupil in the focal plane region can be optically engineered in order to change the simplified view that is now offered. However, for the very restricted case where a LGS is fired along the same directions of the line of sight of the telescope, with an exit pupil located at infinity, one can easily prove that the conjugation of the lower and upper edges of the Sodium layer (regardless of their actual position to a large extent) are embedded one into another as long as the projection telescope is located within the entrance pupil of the telescope. An LGS fired from behind the secondary mirror is not going to be suitable for the following discussion, but as soon as the LGS is projected aside from the main aperture the conjugation of the various points across the Sodium beacon are focussing at independently located positions, without being embedded into adjacent ones.

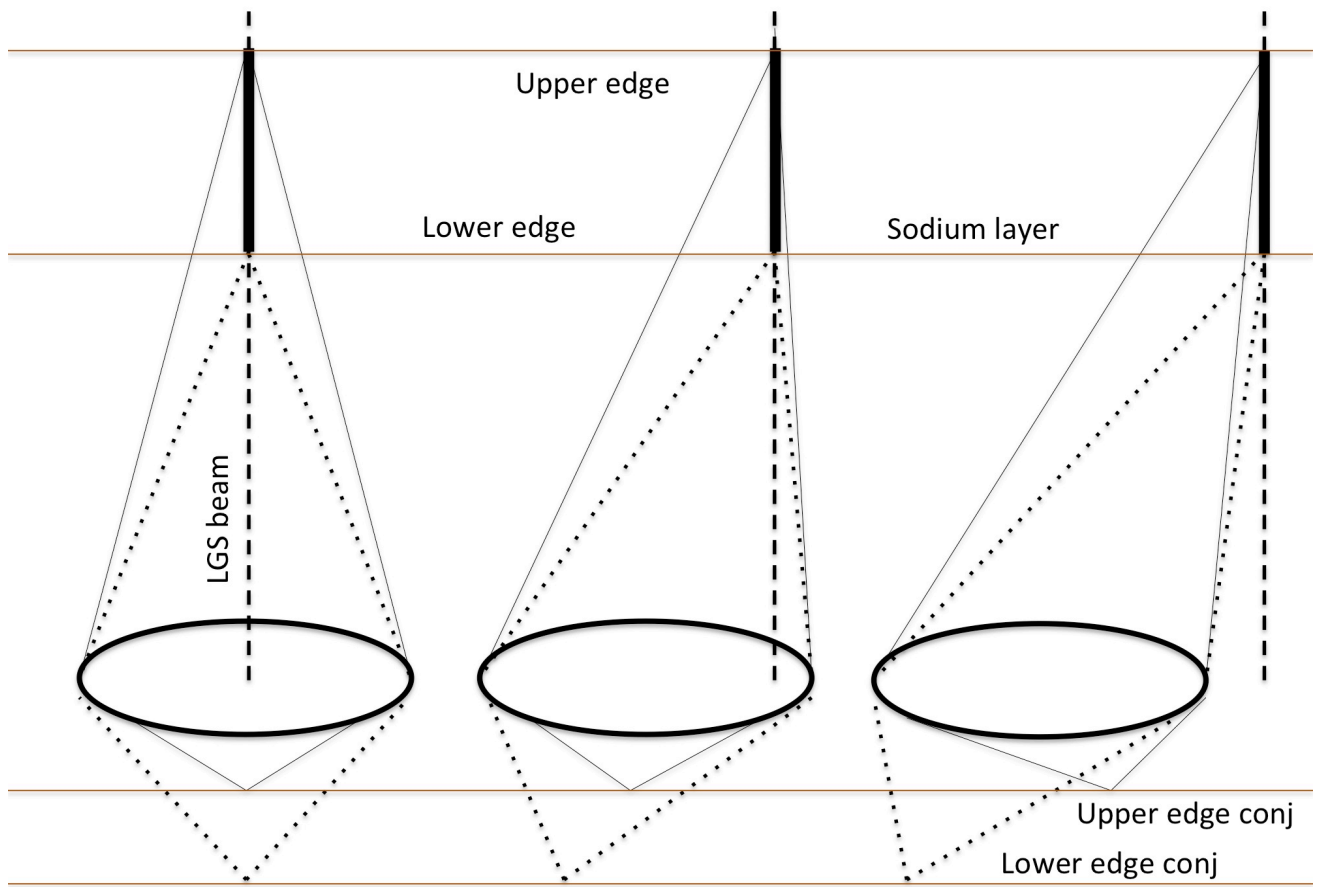

**Figure 1: From left to right: an LGS is projected along the line of sight of the telescope with the projector coaxial, laterally but still in the pupil, and outside the main aperture. The conjugated points are such that the upper edge one is embedded in the beam of the lower one in the first two cases, while it is not in the last one. This is a general situation with the condition described so far, along with the infinite position of the exit pupil plane.**

While in several cases the LGS projection is made, for practical reasons, at a certain distance from the edge of the primary mirror (where, in case of IR optimized telescopes, however, the pupil is not necessarily located) and the exit pupil location can be accommodated with the purpose of adjusting such focus positions one can take advantage of this by splitting in someway the light at this stage and conveying the selected light into a common, or various, pupil reimagers where the pupil is then sampled by a detector with the proper pixel size to match actuators in the DM space. In Fig.2 practical numbering is given for the case of the VLT observing at the local zenith or at an elevation of 45° to give an idea of the size of the volume to be sampled in these cases. We are not going to discuss hereafter the way this splitting is further made, however this can always easily be made by a full reflective system encompassing independent pupil reimagers and detectors, leaving to a further development the option to study solutions allowing a single detector to be used to convey the multiple pupils onto a limited region.

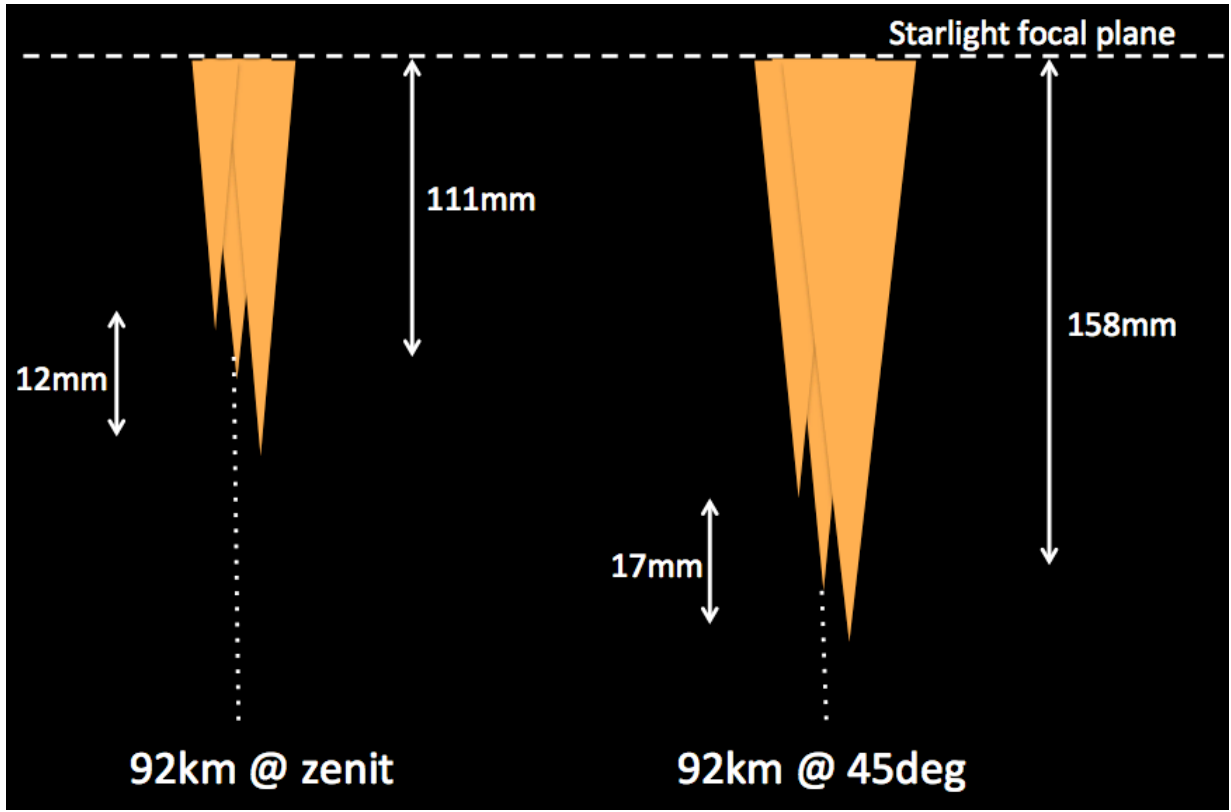

**Figure 2: At a VLT telescope like it is here shown how the optically conjugated points of the inner, upper and central edges of the Sodium layer fall, with respect to the nominal focal plane where natural stars focus. Left: For a zenithal observations. Right: For an observation at 45° of elevation, hence at a 1.414 times longer range and apparent extension of the LGS.**

## 3. TOWARD A 3D SEGMENTATION IN THE FOCAL VOLUME

We now assume that light focusing onto a reimaged 3D copy of the LGS beacon is split in a way that one can select light coming from different portions of it and feeding in this way a collimator illuminating a pupil. In Fig.3 the two most obvious options of how to make this are provided. Basically the idea behind is that the structure of the LGS is here modeled by a uniform elongated source. As the LGS beam is not a geometrical line but it has a significant width it has to be considered, in fact, that even abrupt changes of the Sodium layer density profile will not directly produce, as seen from the telescope subapertures, as sharp features in the reimaged beacon. In fact the width of the upward beam (by diffraction, Fresnel propagation and turbulence encountered during the first path of the launched beam) acts as a sort of spatial filter removing the high spatial frequency content. With the approach described one gives up in toto to use such features to sense the derivative of the wavefront aligned with the direction of the LGS elongation, while the edges of it are instead used for such a purpose. The first approach, hence, foresee the use of a splitting into four different pupils. Two are gathering the light from the edges of the core of the elongated LGS, and the other two collect the light from the opposite edges of the beacon. Thus make a fully independent estimation of the derivative of the wavefront along the elongation and its orthogonal direction. Using such a very crude model it seems like the only "wasting" of light (still using the assumption that the beacon is featureless in its main elongated body) is the lack of the information along the orthogonal to the elongation direction at the level of the two edges. This can be easily accomplished by a further splitting of the light of the edges into two pupils as it is shown in the second option figured out in the right side part of Fig.3. In this way all the six pupils are used to estimate the derivative orthogonally to the elongation, while only four are used to sense the derivative along the elongation itself. However, a more precise scrutiny reveals that the distribution of light into the six portions is not expected to be identical for different subapertures. This is an interesting feature that makes a further distinction in the parallel between this approach and the pyramid one for the natural references. In fact the light on the edges will be more intense, for mere geometrical reasons, for the subapertures that experience a smaller apparent elongation as projected onto a plane orthogonal to the optical axis. So, while the 3D splitting is correct, the way the fluxes are handled takes into account the origin from where a specific ray is coming. This means that also for the estimate of the derivative in the orthogonal direction to the elongation, the light coming from the further splitting of the two edges can be of particular relevance for the subapertures that experience a smaller elongation. Of course this effect is expected to be stronger when the LGS is launched very close to the edge of the pupil while, whenever for practical reasons the projection is made at a significant distance from such an edge, the effect is likely to be of smaller importance, favoring the four pupils arrangement. Simply, in such cases, the difference in elongation between various subapertures becomes less and less an issue.

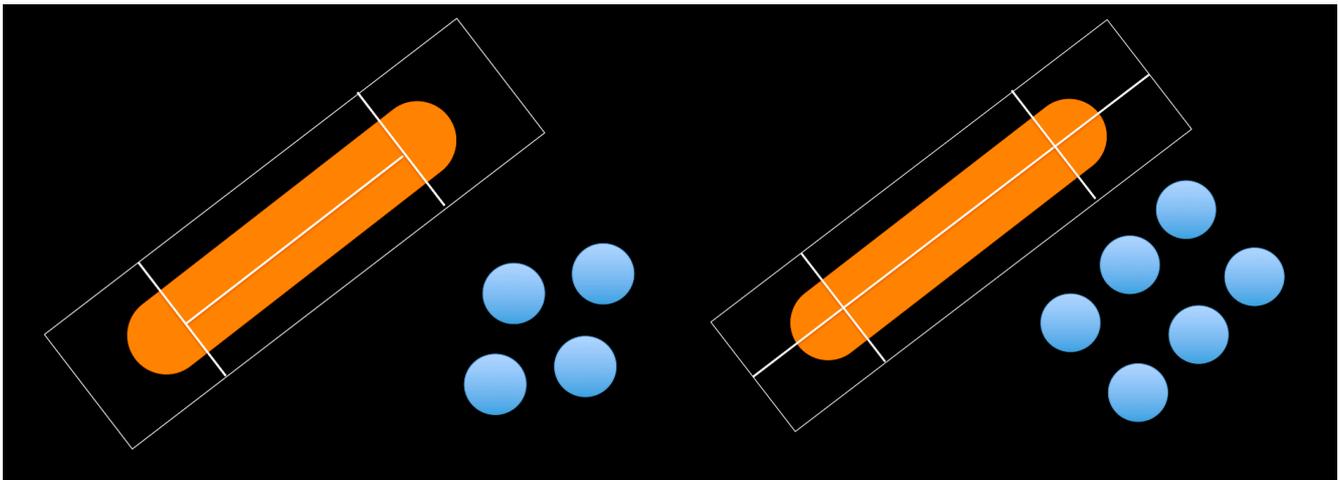

**Figure 3: Left: in this representation a top-view of the LGS as seen projected on the focal volume of the telescope is seen through a subdivision of four different areas. Each of this leading to a pupil plane shown in the lower right inset. Right: The same as the one on the left but with a further subdivision such that the overall number of pupils is six.**

A simple, first order, conjugation point calculation, can give an idea of the physical distances involved in the focal plane position. Recalling the exit pupil at infinity one can see that if the LGS is located at the edge of the entrance pupil the beam focus all lies within the slope of one of the side of the focusing beam itself. As this is usually a very steep angle the plot shown in Fig.4 is anamorphic, or in other words, the scale in the two different axes are significantly different. Some colored dots show off the case of conjugation of points at 50 km, 100 km and 150 km of range. The nominal height of the Sodium layer is 92 km, while its thickness depends upon the definition of where to stop or to apply a threshold for the Sodium column density. Furthermore the actual height or thickness of such a layer is known to have regional or seasonal variations that are not explored here. The largest variation, of course, would simply occurs for perspective change because of the different elevation at which the telescope is continuously moving while observing a given astronomical target. For this region one has to consider ranges that can extend even up to more than 200 km. It is interesting to point that, without considering the Earth curvature, the ratio of the apparent thickness as measured along the line of sight with respect to the distance of the layer remains unchanged. Of course at small elevation angle the equivalent turbulence makes of little practical interest the exploitation of a LGS, or the AO work, almost impossible to attain with decent performances.

One can conceive further elaboration on the scheme and changing placement and arrangement of the pupils in this kind of approach, but these are beyond the limit of this work. However one has to mention that the amount of edges to be used for the derivative estimation along the elongation direction is a matter that can be optimized, and that can be optically tuned with a sort of zoom system changing the plate scale on the splitting device. It is to be recalled that in a four-quadrant system, the scale does not introduce variations on the slope as this is dominated by the ratio of the movement of the beacon with respect to its optical angular size. A shorter focal length will make both a smaller displacement and a corresponding smaller size of the image of the Sodium beacon, leaving invariant the accuracy of the measurement.

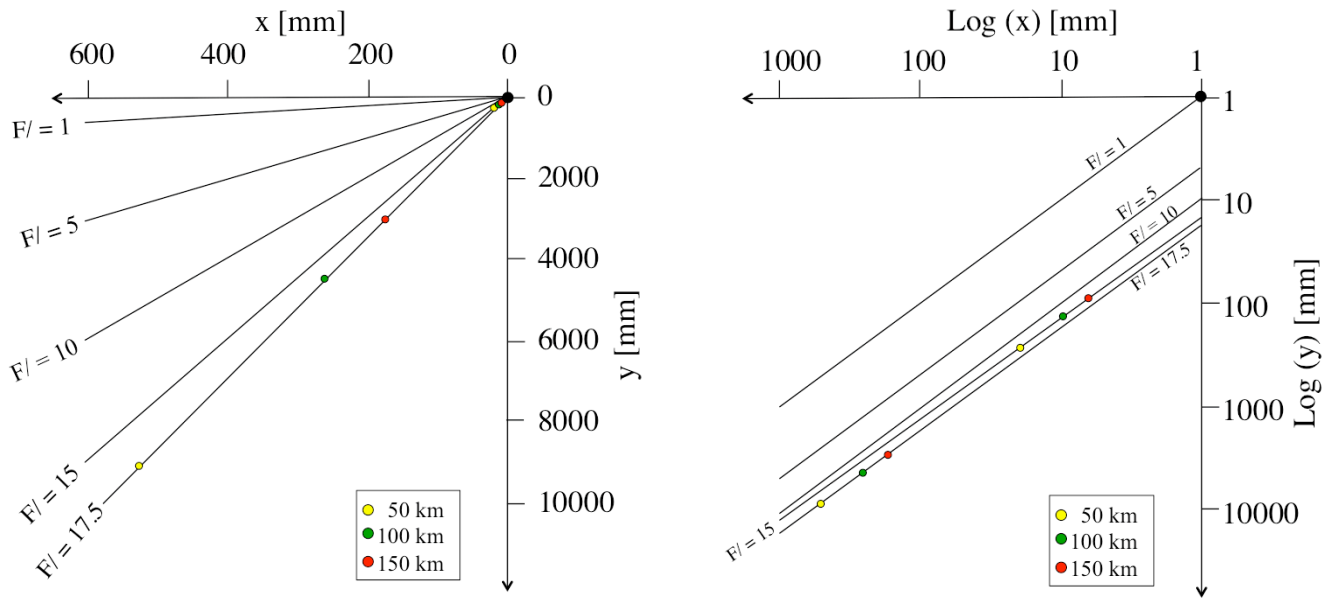

**Figure 4: The conjugation of points at a range of 50 km (yellow), 100 km (green) and 150 km (red) for the VLT and ELT with a reimaging optics working at different focal ratios and for (left) a linear anamorphic scale and (right) a log-log anamorphic scale.**

## 4. CONCLUSIONS

A first-order description of a device that promises to produce a detector occupancy similar to a pyramid one, but tuned to a LGS beacon has been shown in a conceptual fashion. This device promises to exhibit an efficient use of a CCD-like detector with an optimum use of the LGS beacon in its 6 pupil version, without taking into account and taking advantage of the features along the elongated image of the Sodium beacon. Further work is required on the optomechanical side to finalize a solution that is both compact and easy to produce, and to actually estimate the gain with respect to other more classical approach. The proposed system, however, looks like a very efficient solution for a non-moving part LGS wavefront sensor that does not play any constraints on the laser format as it works with CW as well as any pulsed format lasers.